\documentclass[amsmath,amssymb,groupedaddress]{revtex4}
\usepackage{graphicx}
\usepackage{epstopdf}
\usepackage{xcolor}
\usepackage{amsmath,amsthm,amssymb}
\newcommand{\be}{\begin{equation}}
\newcommand{\ee}{\end{equation}}
\newcommand{\ba}{\begin{eqnarray}}
\newcommand{\ea}{\end{eqnarray}}
\newcommand{\baa}{\begin{eqnarray}}
\newcommand{\eaa}{\end{eqnarray}}
\newcommand{\ed}{\end{document}}
\newcommand{\lab}[1]{\label{#1}}
\newcommand{\re}[1]{(\ref{#1})}

\begin{document}
\title{Fokker-Planck equation on metric graphs}
\author{J. Matrasulov$^{a}$, K. Sabirov$^c$}

\affiliation{${^a}$Yeoju Technical Institute in Tashkent, 156 Usman Nasyr Street, 100121 Tashkent, Uzbekistan\\
$^c$Tashkent University of  Information Technology, Amir Temur
Avenue 108, Tashkent 100200, Uzbekistan}
\begin{abstract}
 We consider the Fokker-Planck equation on metric graphs.
 Vertex boundary conditions are imposed in the form of weight continuity and the  probability current conservation. Exact solution of the Fokker-Planck equation on star, tree and loop graphs is obtained. Applications of the model to Brownian motion in networks and other problems are briefly discussed.
\end{abstract}

\maketitle
\section{Introduction}
The Fokker-Planck equation is an evolution equation that  governs an important class of Markov processes and that is used for modeling wide class of stochastic processes in statistical mechanics \cite{FPEB1}-\cite{Sorokin}. It describes   the time evolution of the probability density function of the position of a particle but not the velocity of a particle under the influence of drag forces and random forces. So far,  the Fokker-Planck equation found numerous  applications in plasma physics \cite{Plasma,FPE Plasma 1}, particle and beam physics \cite{FPE particle}, thermodynamics \cite{Thermo,FPE Thermodynamics}, condensed matter \cite{FPE Condensed}, real gases \cite{FPE Gas,FPE Gas1}, fluid dynamics \cite{FPE Fluid}, neural networks \cite{Neural,Neural2,Neural learning,Neural networks,Neural networks1}, traffic modelling \cite{FPE Traffic,FPE Traffic 1,FPE Traffic 2} and even in socio- and econo-physics \cite{FPE Econophysics,FPE Econo,FPE Econo1,FPE Econo2,FPE Econosocio}. Different aspects of mathematical properties \cite{Nagafi} of the Fokker-Planck equation (FPE) have been widely studied and various schemes have been proposed \cite{FPE Numer} for its numerical solution. In this paper we address the problem of Fokker-Planck equation in networks. These latter are the quasi-one-dimensional domains having branched structure, which can be modelled in terms of so-called metric graphs, i.e., the branched wires connected to each other according to the rule, called topology of a graph. We obtain exact solution of the FPE on three types of graphs. However, the approach is applicable for wide class of network topologies, except, or course, very complicated general graphs (e.g., for directed graphs). The topic of partial differential equations (PDE) on graphs has attracted much attention recently (see, e.g., Refs.\cite{Uzy2,Uzy3,Uzy4,Mugnolo,Grisha,Exner15,JRY,EPL,PRE,Mashrab,PT,PRE1,PLA,JPA,JCP,EPL11,PRE3} for review). The motivation for the study of the evolution equations on
networks comes from the fact that the dynamics
of waves in networks is richer than that in a line, i.e. unbranched domains. 
In addition, by choosing the proper network architecture (topology),
one can achieve the required wave propagation/transport 
regime, i.e., a tunable  wave dynamics. Also, in many practically important
applications (e.g., traffic flow, brownian motion, capillary or microtube flows, etc.) one deals with branched structures rather than lines. In
such structures, the branching architecture can be used for the controlling the the dynamics. The dynamics of quasiparticles and waves strongly depend on the network architecture. More complicated network topology provides more richer  dynamics. In particular, due to the transmission and/or back-scattering at the nodes the wave or particle can achieve diffusive or ballistic motion, as well as change the direction of motion.  Thus using FPE on networks by modeling these latter in terms of metric graphs provides powerful tool for the problem of tunable stochastic evolution in networks.  This paper is organized as follows. In the next section we present brief description of the Fokker-Planck equation on a line. Section III presents formulation of the problem on FPE on a star graph, its analytical and numerical solutions. In section IV we extend the model to other graphs topologies, such as tree and H-graphs. Finally, section V presents some concluding remarks.

\section{Fokker-Planck equation on a real line}
Before proceeding to the Fokker-Planck equation on a graph, we briefly recall solution of  Fokker-Planck equation (FPE) on a real line following the Ref. \cite{FPEHR1}.
General form of Fokker-Planck equation can be written as

\be 
\frac{\partial\textit{f}}{\partial
t}=\gamma\frac{\partial}{\partial
x}\Bigl( \frac{dV(x)}{dx}\textit{f}\Bigr)+D\frac{\partial^2}{\partial
x^2}\textit{f}.\lab{fpe0} 
\ee
Here we will consider FPE for the diffusive harmonic
oscillator which is given as
 \cite{FPEB1} \be \frac{\partial\textit{f}}{\partial
t}=\gamma\frac{\partial}{\partial
x}(x\textit{f})+D\frac{\partial^2}{\partial
x^2}\textit{f},\lab{fpe1} \ee where $\textit{f}(x,t)$ is a
probability density, $\gamma$ is the drift velocity (describes level of drifting, e.g., of Brownian particle) and $D=const$ is the diffusion coefficient (in general $D$
can be  function of the time and the space, but here, for simplicity for consider it as constant).

For the initial conditions imposed as \be
\textit{f}(x,0)=\delta(x-x_0), \ee the solution of Eq.\re{fpe1} can
be written as \cite{FPEHR1}
\be
\textit{f}(x_0,x,t)=\sqrt{\frac{\gamma}{2\pi
D(1-e^{-2\gamma t})}}\exp\Biggl[-\frac{\gamma(x-e^{-\gamma
t}x_0)^2}{2D(1-e^{-2\gamma t})}\Biggr]. \ee The steady state
solution, i.e. when $\frac{\partial \textit{f}}{\partial t}=0$, is: \be
\textit{f}=\sqrt{\frac{\gamma}{2\pi D}}\exp\Biggl[-\frac{\gamma
x^2}{2D}\Biggr]. \lab{solu}\ee.

Finally, we note that the Fokker-Planck equation \re{fpe1} approves the following
conservation law: \be
\frac{d}{dt}\int\limits_{-\infty}^{+\infty}\textit{f}(x,t)dx=0.
\lab{cons1} \ee

\begin{figure}[h!]
\includegraphics[width=8cm]{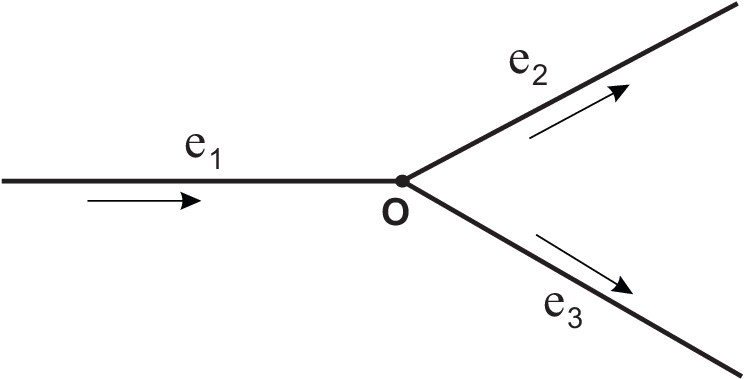}
\caption{Simplest star graph}
\label{star}
\end{figure}
\section{Fokker-Planck equation on a simple star graph}
Here we address the problem of Fokker-Planck equation on graphs considering first simplest graph topology in the form of Y-junction, basic star graph. Such graph is the basic unit that allows to construct arbitrary or most of the graph structures (architectures).   
We note that the problem evolution equation on graphs has become separate topic in mathematical and theoretical physics during past decade. The graphs are defined as the quasi-one-dimensional domains consisting of bonds connected to each other according to the rule, which is called topology of a graph. An evolution equation on graphs is written on each bond and each equation related to other through the vertex boundary conditions, which are imposed at the nodes of a graph. An effective approach for solving evolution equations on graphs was proposed in \cite{Zarif} for nonlinear Schrodinger equation and applied later to sine-Gordon \cite{EPL11} and nonlinear Dirac \cite{PRE1} equations. The method is based on using the solution of an evolution equation on a line to construct solution on a graph by fulfilling vertex boundary conditions. Solutions constructed in such a way fulfill the vertex boundary conditions under certain constraints which are given in terms of task parameters. Here we will apply this approach to Fokker-Planck equation on a metric star graph, which is presented in Fig. 1. On each bond
$e_j$, of the graph we assign a coordinate $x_j$, which indicates the position
along the bond: for bond $e_1$ it is $x_1\in (-\infty,0]$ and for $e_{1,2}$
they are $x_{2,3}\in [0,+\infty)$.
In what follows,  we will use the notation $f_j(x)$ for $f_j(x_j)$ and it is understood that $x$ implies the coordinate on the bond $j$ to which the component $f_j$ refers.
On such graph FPE can be written as
 \begin{equation}
\frac{\partial\textit{f}_j}{\partial
t}=\gamma_j\frac{\partial}{\partial
x}(x\textit{f}_j)+D_j\frac{\partial^2}{\partial
x^2}\textit{f}_j,\;\; j=1,2,3. \label{fpe2}\end{equation}
Here $j$ is the bond number, $D_j$ is the diffusion coefficient for $j\;$ th bond, $f_j$ is the probability density for $j\;$th bond. To solve Eq.\re{fpe2}.
one needs the boundary conditions at the vertex (branching point). Here we impose weight continuity for $f_j$ as
\begin{equation}
\sqrt{\frac{D_1}{\gamma_1}}f_1(0,t) =\sqrt{\frac{D_2}{\gamma_2}} f_2(0,t) =\sqrt{\frac{D_3}{\gamma_3}} f_3(0,t),
\end{equation}
and the boundary conditions following from the current conservation, which are given by Eq.(\ref{cons1})
\begin{equation}
D_1\frac{\partial f_1}{\partial x}=D_2\frac{\partial f_2}{\partial x}+D_3\frac{\partial f_3}{\partial x}.
\lab{vbc01}
\end{equation}

To complete formulation of the task, we need to impose also initial conditions. Here we choose solutions for the initial conditions imposed as 
\begin{equation}
f_j(x,0)=\delta(x-x_{0j}).\label{initial1} 
\end{equation}
Such initial condition is often used in many exactly solvable cases e.g., for Brownian motion (see, Refs.\cite{FPEB1}-\cite{Bolivar} for review of solutions of FPE). For the graph, presented in Fig.1, the peak of the delta-function always belongs to the graph bond, as the length of the letter is semi-infinite.
Solution of Eq.(\ref{fpe2}) fulfilling the above vertex boundary conditions and the initial condition (\ref{initial1}) can be written as
\begin{equation}
\textit{f}_j(x_{0j},x,t)=\sqrt{\frac{\gamma_j}{2\pi
D_j(1-e^{-2\gamma_j t})}}\exp\Biggl[-\frac{\gamma_j(x-e^{-\gamma_j
t}x_{0j})^2}{2D_j(1-e^{-2\gamma_j t})}\Biggr]. \label{sol01}
\end{equation}
It is important to note that solution given by Eq.(\ref{sol01}) fulfill the vertex boundary conditions provided the following sum rule is valid:
\begin{equation}
\frac{1}{\sqrt{D_1}}=\frac{1}{\sqrt{D_2}}+\frac{1}{\sqrt{D_3}},
\label{sr1}
\end{equation}
with $\gamma_1=\gamma_2=\gamma_3=\gamma,\,x_{0j}=\pm \sqrt{D_j}$, i.e., the drift velocities are chosen as the same for all bonds.
In other words, solution given by Eq.\re{sol01} obeys Eq.(\ref{fpe2}) only in case, if the above sum rule is fulfilled, if sum rule is broken, one needs to solve Eq.(\ref{fpe2}) numerically. Thus,  unlike the FPE on a real line, for the case of graphs, the analytical results, i.e. the exact solution of the problem can be obtained for the special case, given in terms constraints in Eq.\re{sr1}. This is the "cost", one needs to pay to obtained exact solution of the problem on a graph. In Fig.2 solution of Eq.(\ref{fpe2}) is plotted for the case, when sum rule given by Eq.(\ref{sr1}) is fulfilled (for different time moments, $t=150\;$ $t=700\;$ and $t=1000$). Fig.3 presents numerically obtained  plots of the solution, $f_j(x)$ for the case, when the sum rule (\ref{sr1}) is broken. Crank-Nicolson discretization scheme implemented for FPE in the Ref. \cite{CN} is used for numerical solution of Eq.(\ref{fpe2}) with the vertex boundary and initial conditions given by Eqs. \re{vbc01} and (\ref{initial1}), respectively.   Certain qualitative and quantitative difference in the profiles for two cases can be observed from the plots. Fig. 4 presents plots of the average coordinate computed using the probability density, $f(x)$. It is the main characteristics, e.g., Brownian motion. The plot is obtained for the case, when the initial condition is imposed on the first bond. Upon growing in the first bond, the average velocity starts to decay.  We note that the condition (sum rule) in Eq.(\ref{sr1}) ensures existence of the exact solution of the FPE on metric star graph and its uniqueness for the initial condition \re{initial1}.
\begin{figure}[h!]
\includegraphics[width=15cm]{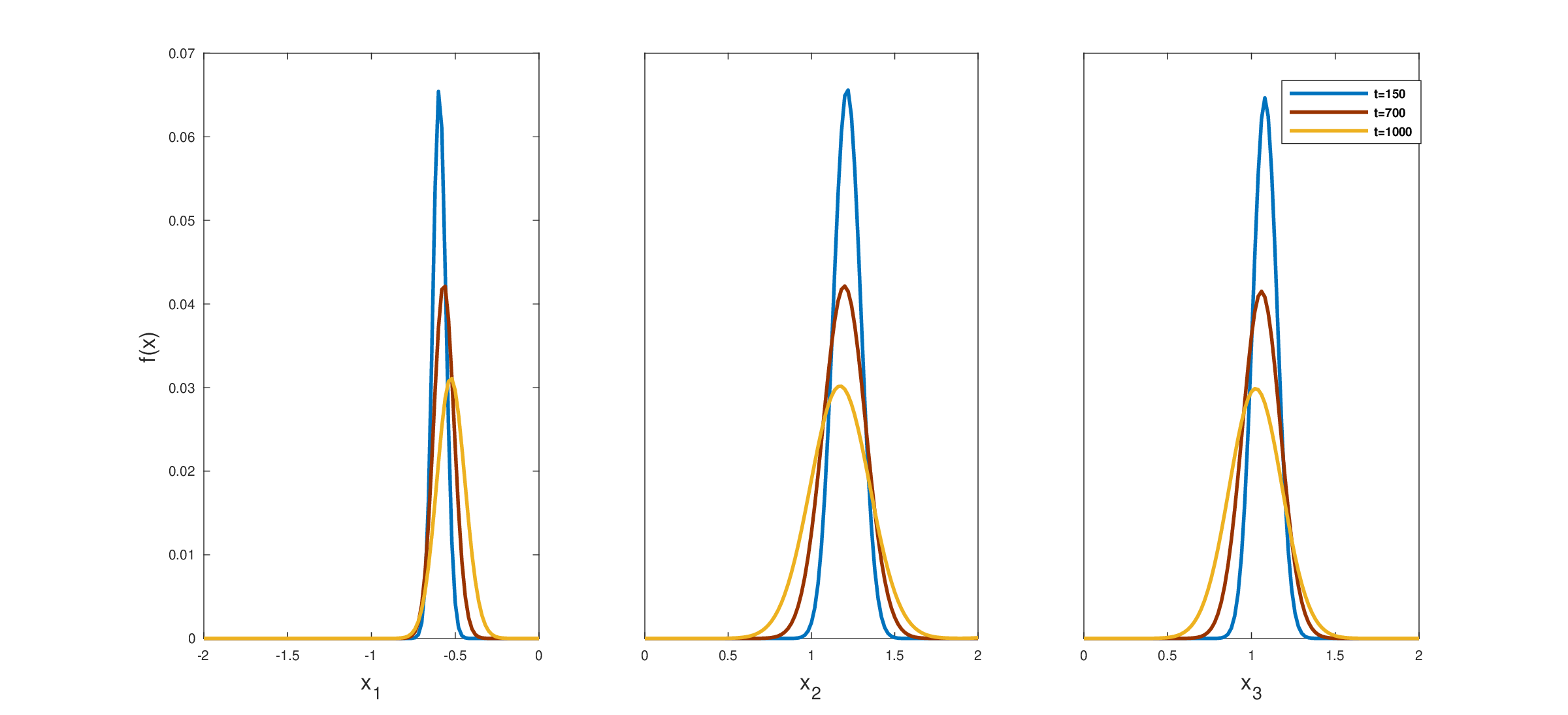}
\caption{Probability density, $f(x)$ for star graph, obtained by solving Eq.\re{fpe2} for the case, when sum rule \ref{sr1} is obeyed, with $x_{01}=0.7$, $x_{01}=0.3$, $x_{01}=0.5$ $\gamma_1=\gamma_2=\gamma_3=1.5$ and  $D_2=0.9$, $D_3=0.7$.}
\label{fig1}
\end{figure}

\begin{figure}[h!]
\includegraphics[width=15cm]{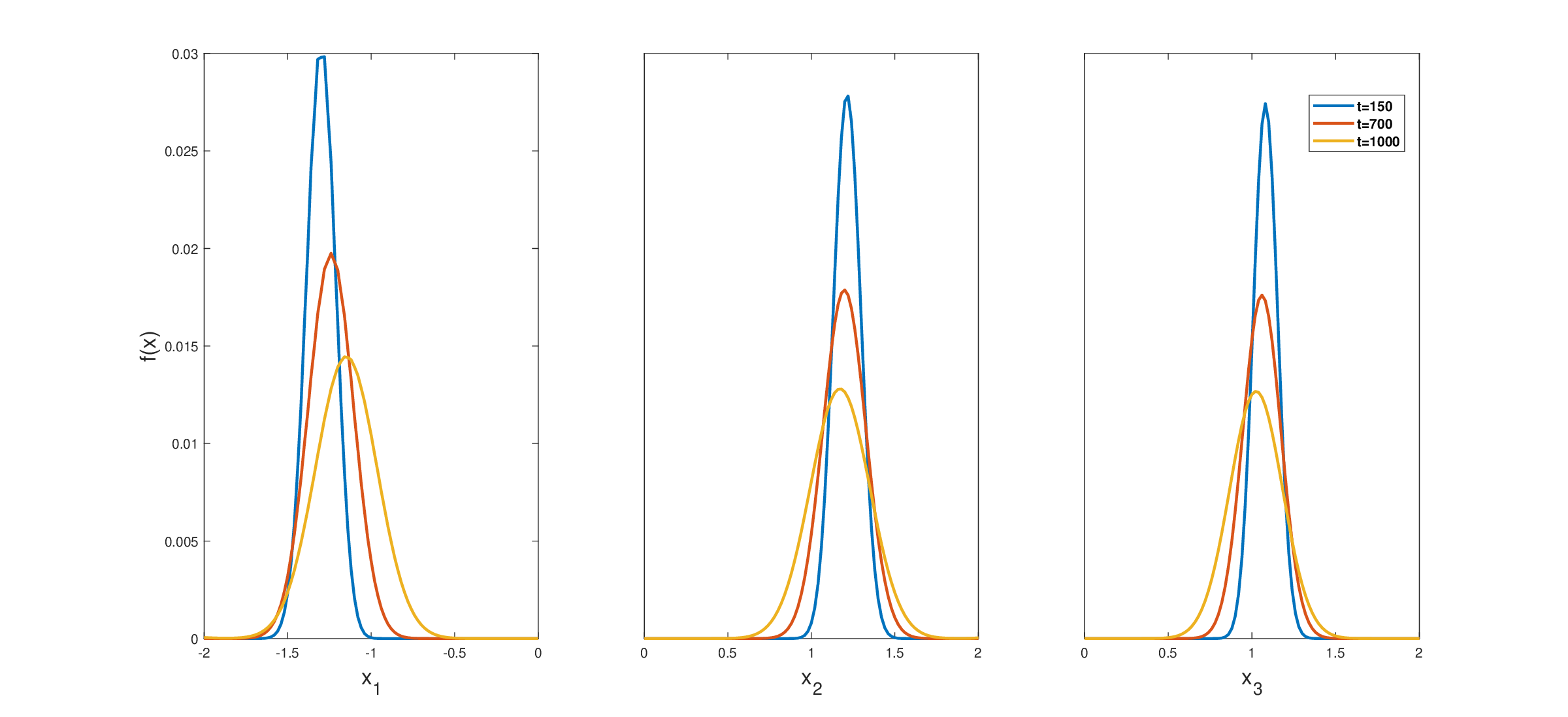}
\caption{Probability density, $f(x)$ for star graph, obtained by solving Eq.\re{fpe2} for the case, when sum rule \re{sr1} is broken, with  $x_{01}=0.7$, $x_{01}=0.3$, $x_{01}=0.5$ $\gamma_1=\gamma_2=\gamma_3=1.5$ and $D_1=0.5$ $D_2=0.9$, $D_3=0.7$.}
\label{fig2}
\end{figure}

\begin{figure}[h!]
\includegraphics[width=15cm]{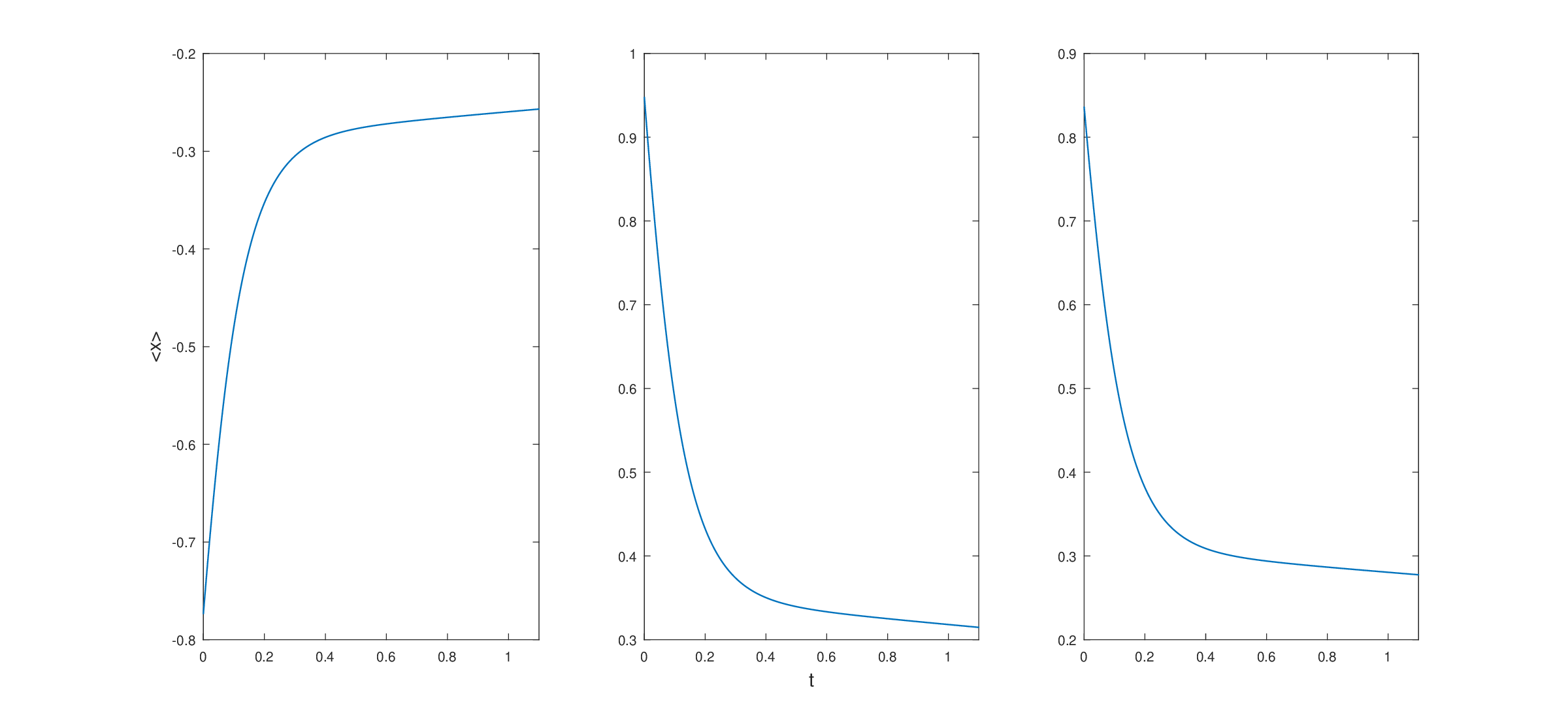}
\caption{Average coordinate of a Brownian particle on a star graph, for $\gamma_1=\gamma_2=\gamma_3=1.5$ and  $D_1=1.2$, $D_2=0.9$, $D_3=0.7$.}
\label{fig3}
\end{figure}
\newpage
\section{Extension to other graphs}
Although the treatment of the previous section was done for simplest graph, the above approach for solving FPE on graphs can be directly extended to other graph topologies. Here we demonstrate this for H-graph (see, Fig.5) and for tree graph presented in Fig.6.
For H-graph the Fokker-Planck equation can be written (on each bond) as
 \begin{equation}
\frac{\partial\textit{f}_j}{\partial
t}=\gamma_j\frac{\partial}{\partial
x}(x\textit{f}_j)+D_j\frac{\partial^2}{\partial
x^2}\textit{f}_j,\;\; j=1,2,3, 4,5. \label{fpe3}\end{equation}.

First set of the vertex boundary conditions at the nodes $O_1$ and $O_2$ can be imposed as weight continuity:
$$
\sqrt{\frac{D_1}{\gamma_1}}f_1(0,t) = \sqrt{\frac{D_2}{\gamma_2}}f_2(0,t) = \sqrt{\frac{D_3}{\gamma_3}}f_3(0,t),
$$
\begin{equation}
\sqrt{\frac{D_3}{\gamma_3}}f_3(L_3,t) = \sqrt{\frac{D_4}{\gamma_4}}f_4(L_4,t) = \sqrt{\frac{D_5}{\gamma_5}}f_5(L_5,t),
\end{equation}
while, second set can be derived from the current conservation law:
\begin{equation} 
D_1\frac{d\textit{f}_1(x,t)}{dx}|_{x=0}+D_2\frac{d\textit{f}_2(x,t)}{dx}|_{x=0}=D_3\frac{d\textit{f}_3(x,t)}{dx}|_{x=0},
\end{equation}
\begin{equation} 
D_3\frac{d\textit{f}_3(x,t)}{dx}|_{x=L_3}=D_4\frac{d\textit{f}_4(x,t)}{dx}|_{x=L_4}+D_5\frac{d\textit{f}_5(x,t)}{dx}|_{x=L_5}.
\end{equation}
Furthermore, we require that solution of FPE on a line, given by Eq.\ref{solu} is made the scaling $x=\sqrt{D_j}y$  on bonds $3,4,5$ with $L_{3,4,5}=\sqrt{D_{3,4,5}}l$  and should fulfill these vertex boundary conditions. It is easy to see that this is possible, if the parameters $D_j$  fulfill the following sum rules:
\be
\frac{1}{\sqrt{D_1}}+\frac{1}{\sqrt{D_2}}=\frac{1}{\sqrt{D_3}}
\lab{sr01},
\ee
\be 
\frac{1}{\sqrt{D_3}}=\frac{1}{\sqrt{D_4}}+\frac{1}{\sqrt{D_5}},
\lab{sr02}
\ee
where we assumed the following relations between other parameters:\\
$\gamma_1=\gamma_2=\gamma_3=\gamma_4=\gamma_5=\gamma$, $x_{0j}=\pm\sqrt{D_j}$ .
Provided the sum rules (constraints) given by Eqs.\re{sr01} and \re{sr02} are fulfilled, the solution of Eq.\re{fpe2} can be written as
\be
\textit{f}_j(x_{0j},x,t)=\sqrt{\frac{\gamma_j}{2\pi
D_j(1-e^{-2\omega_j t})}}\exp\Biggl[-\frac{\gamma_j(x-e^{-\gamma_j
t}x_{0j})^2}{2D_j(1-e^{-2\gamma_j t})}\Biggr],\; j=1,2,3,4,5. 
\lab{sol03}
\ee
 Similarly to the above, one can obtain solution of the Fokker-Planck equation for the tree graph, presented in Fig. 5. On each bond of tree graph we have Eq.(\ref{fpe2}).

\begin{figure}[h!]
\includegraphics[width=8cm]{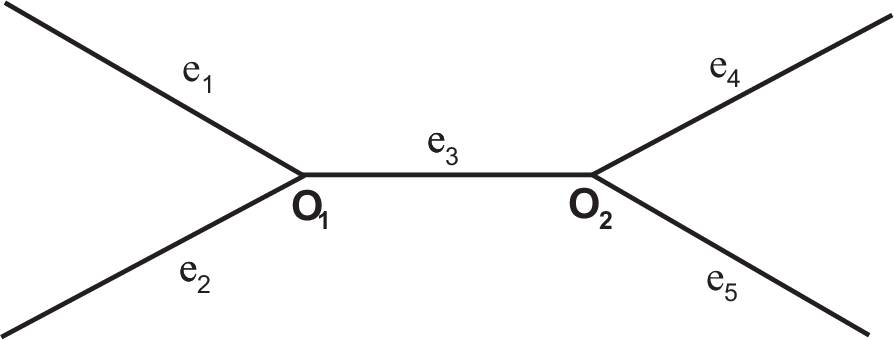}
\caption{H graph}
\label{Hgraph}
\end{figure}
 
The vertex boundary conditions are given at the nodes $O_1,\;$ $O_{11},\;$ and $O_{12}$:
$$
\sqrt{\frac{D_1}{\gamma_1}}f_1(0,t) = \sqrt{\frac{D_{11}}{\gamma_{11}}}f_{11}(0,t) = \sqrt{\frac{D_{12}}{\gamma_{12}}}f_{12}(0,t),
$$
\begin{equation}
\sqrt{\frac{D_{1i}}{\gamma_{1i}}}f_{1i}(L_{1i},t) = \sqrt{\frac{D_{1i1}}{\gamma_{1i1}}}f_{1i1}(L_{1i1},t) = \sqrt{\frac{D_{1i2}}{\gamma_{1i2}}}f_{1i2}(L_{1i2},t),
\end{equation}
and
$$
D_1\frac{d\textit{f}_1(x,t)}{dx}|_{x=0}=D_{11}\frac{d\textit{f}_{11}(x,t)}{dx}|_{x=0}+D_{12}\frac{d\textit{f}_{12}(x,t)}{dx}|_{x=0},
$$
\begin{equation} 
D_{1i}\frac{d\textit{f}_{1i}(x,t)}{dx}|_{x=L_{1i}}=D_{1i1}\frac{d\textit{f}_{1i1}(x,t)}{dx}|_{x=L_{1i1}}+D_{1i2}\frac{d\textit{f}_{1i2}(x,t)}{dx}|_{x=L_{1i2}}.
\end{equation}
These boundary conditions will be fulfilled by the solution of FPE on a line, provided the following sum rules hold true:
\begin{equation}
\frac{1}{\sqrt{D_1}}=\frac{1}{\sqrt{D_{11}}}+\frac{1}{\sqrt{D_{12}}},
\end{equation}
\begin{equation}
\frac{1}{\sqrt{D_{1i}}}=\frac{1}{\sqrt{D_{1i1}}}+\frac{1}{\sqrt{D_{1i2}}},
\end{equation}
where the following relations are assumed: $\gamma_1=\gamma_{1i}=\gamma_{1ij}=\gamma,\,x_{01}=\pm\sqrt{D_1},\,x_{01i}=\pm\sqrt{D_{1i}},\,x_{01ij}=D_{1ij},\,i,j=1,2$.
Then the solution of Eq.(\ref{fpe3}) for the tree graph presented in Fig.5 can be written as
\begin{equation}
\textit{f}_e(x_{0e},x,t)=\sqrt{\frac{\gamma_e}{2\pi
D_e(1-e^{-2\gamma_e t})}}\exp\Biggl[-\frac{\gamma_e(x-e^{-\gamma_e
t}x_{0e})^2}{2D_e(1-e^{-2\gamma_e t})}\Biggr]. 
\label{sol04}
\end{equation}

\begin{figure}[h!]
\includegraphics[width=8cm]{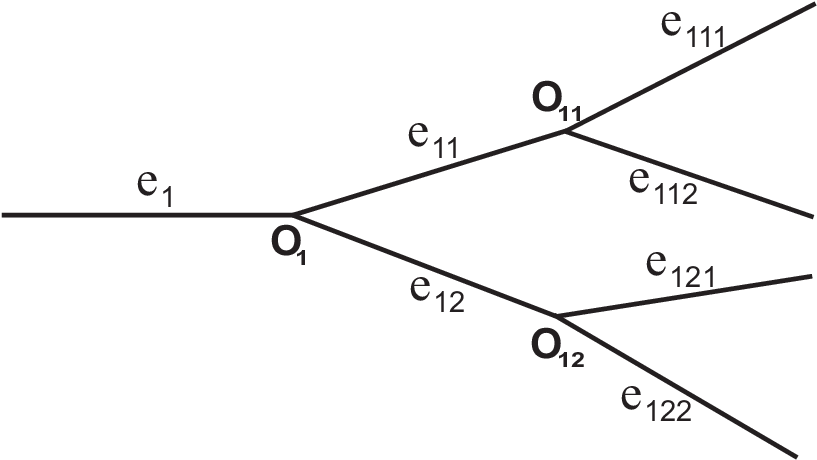}
\caption{A sketch for tree graph}
\label{tree graph}
\end{figure}
We note that in case of H- and tree-graphs the initial conditions should be imposed 
on semi-infinite bonds, where initial incoming wave appears. This ensures location of
the delta-peaks on the graph bonds. Although we have shown extension of the approach of section II to H- and tree graphs, it can be applied for arbitrary graph, provided graph consists of sub-graph with finite lengths and at least three outgoing, semi-infinite bonds. However, we note that the  closed graphs, e.g., graphs with cycles do not approve exact solution within our approach. The above treatment concerns Fokker-Planck equation with simple, linear potential, more complicated potential also can be considered within our approach. In particular, exact solution of FPE on a graph can be obtained provided the problem with a complex potential approves exact solution on a line. Of course, one always can solve the problem numerically for arbitrary (or, at least for wide class) of potential.

\section{Conclusions}
In this paper we studied Fokker-Planck equation on networks by modeling these latter in terms of metric graphs. Exact analytical solutions are obtained for the vertex boundary conditions given in the form of weight continuity and Kirchhoff's rules. Constraints providing validity of such solutions are derived in terms of simple sum rules for the values of the diffusion coefficient $D_j$. Numerical solutions of FPE are obtained for the case, when the sum rules are broken. The above developed approach can be extended for solving the Fokker-Planck equation on arbitrary graph. The results obtained can be directly applied for modeling Brownian motion in branched structures and networks. Also, such problems as traffic modeling, signal propagation in neural networks and plasma dynamics in branched waveguides can be modeled using the above developed approach. Finally, we note that the above treatment deals with the "open" graphs, i.e. the graphs having at least three semi-infinite bonds. However, the approach used in the paper can be applied also for closed graphs, with finite bonds. To construct solution of the problem for closed graphs one needs to use solution of FPE on a finite interval. Solution of this task is on progress now and should be subject for forthcoming paper.

\section{Acknowledgements} This work is supported by a grant  of the Ministry for Innovation Development of Uzbekistan (Ref. No. FZ-20200928103).

\end{document}